\documentclass[preprint,showpacs,preprintnumbers,amsmath,amssymb]{revtex4}

\usepackage{graphicx}% Include figure files
\usepackage{dcolumn}% Align table columns on decimal point
\usepackage{bm}% bold math

\begin{document}

%\preprint{APS/123-QED}

\title{Equilibrium properties of highly asymmetric star-polymer mixtures}% Force line breaks with \\

\author{Christian Mayer}
%\email{mayer@thphy.uni-duesseldorf.de}
\author{Christos N. Likos}
\author{Hartmut L\"owen}
 \affiliation{Institut f{\"u}r Theoretische Physik II,
              Heinrich-Heine-Universit{\"a}t D{\"u}sseldorf, 
              Universit{\"a}tsstra{\ss}e 1, D-40225 D{\"u}sseldorf, Germany
}

\date{{\bf \today}, submitted to {\sl Physical Review E}}

\begin{abstract}
We employ effective interaction potentials to study the equilibrium
structure and phase behavior of highly asymmetric mixtures of star
polymers. We consider in particular the influence of the addition of
a component with a small number of arms and a small size on a 
concentrated solution of large stars with a high functionality. 
By employing liquid integral equation theories we examine the evolution
of the correlation functions of the big stars upon addition of the
small ones, finding a loss of structure that can be attributed to 
a weakening of the repulsions between the large stars due to the presence
of the small ones. We analyze this phenomenon be means of a generalized
depletion mechanism which is supported by computer simulations. 
By applying thermodynamic perturbation theory we draw the phase
diagram of the asymmetric mixture, finding that the addition of small
stars melts the crystal formed by the big ones. A systematic comparison
between the two- and effective one-component descriptions of the
mixture that corroborates the reliability of the
generalized depletion picture is also carried out.
\end{abstract}


\maketitle

\section{Introduction}

Mixtures whose constituent particles show a high asymmetry in sizes
are quite common in soft matter physics. As a matter of fact, all
soft-matter systems are at least two-component mixtures, as they
are typically suspensions or dispersions of mesoscopically-sized
colloidal particles in a microscopic solvent. For many practical purposes,
though, it suffices to model the solvent as a continuous medium and
then an effective, one-component description of the suspended colloidal
particles is sufficient. The phenomenology is much richer when more
than one species of colloids is dispersed in the solvent and also there
the asymmetry in the sizes of the two kinds of colloidal particles can
be much higher than the one encountered in atomic or molecular fluids. 
In the recent past, a great deal of attention has been paid to the
investigation of model colloid-polymer mixtures, in which the two 
species are hard colloidal spheres and soft, flexible polymer 
chains \cite{poon:review}.
The bulk of the theoretical analysis of such systems is carried out
within the framework of the Asakura-Oosawa (AO)
model \cite{ao:54, ao:56, vrij:76}, in which the
polymers are modeled as ideal, interpenetrating spheres that experience
a hard repulsion towards the colloids. Another popular system that
has attracted a lot of attention recently are binary hard-sphere
mixtures of various size ratios \cite{biben91, biben91a, md1, md2, md3}.
In both of those cases, attention is
usually focussed on the influence of the smaller component on the
structural and phase behavior of the larger one \cite{fuchs:review}.
Demixing phase
transitions and their competition to the crystallization transition
of the large hard spheres have been an issue of intensive investigations
in the past \cite{ilett:95, louis:99, dijkstra:99, matthias:prl:00, fuchs:00, matthias:jpcm:02, matthias:jcp:02, louis:prl:02}
with current research steering in the direction
of the study of interfacial and wetting properties of such 
mixtures \cite{evans:04,brader:00, brader:01, brader:02, dijkstra:02, wijting:03, moncho:03, wessels:04},
as well as the influence of the additives on the vitrification 
transition of the hard spheres \cite{bergenholtz:99, johner:01, foffi:02, puertas:03, bergenholtz:03, pham:04, zacca:04}.

A convenient concept that has helped shed light into the phenomenology
of such asymmetric mixtures is that of the effective, {\it depletion
interaction} between the hard spheres, which is mediated by the
smaller component \cite{likos01, louis:jpcm:01}.
In the case of the AO-mixture, the depletion interaction
is purely attractive and has the range of the size of the added
polymer. For binary hard-sphere mixtures, the effective depletion
potential displays oscillatory behavior due to 
correlation effects \cite{goetz:99, roth:00}.
Interpolating between the two extremes of ideal and hard additives
are star polymers of varying functionality, whose depleting effects on
hard spheres have been investigated both 
theoretically \cite{dzubiella02, dzubiella02a} and 
experimentally \cite{joe:pre}.

The notion of depletion is almost exclusively invoked whenever the
large particles are hard colloids. Nevertheless, it can be expanded
in its interpretation to account for the modification of the properties
of the large particles in the presence of smaller ones also for arbitrary
kinds of interactions between the constituent particles. There is
relatively little done in this direction, however, with the 
exception of the derivation of effective potentials in 
Yukawa mixtures \cite{roth:02}
and in mixtures of star polymers and linear chains \cite{stiakakis02}.
In the last case,
it has been shown that the depletion mechanism of the chains on the
stars can account for the experimentally observed melting of the 
star-polymer gel upon addition of linear polymer. In this paper, we
turn our attention to two-component mixtures in which all particle
species interact by means of soft potentials and, in particular,
to mixtures of two kinds of star polymers: large ones with a high
number of arms and small ones will a low arm number. All species
interact via logarithmic-Yukawa pair potentials. We find that in this
case the depletion mechanism of the small stars on the big ones
has the effect of reducing the repulsive potential between the latter 
and thus it brings about a melting of the colloidal crystal formed
by the large stars. Concomitant to this effect is a partial loss
of correlations between the centers of the big stars, manifested in
a drastic lowering of the peak height of their partial structure
factor. Upon addition of a sufficiently large quantity of depleting
agents, even an effective {\it attraction} between the large stars
shows up, resulting in a demixing spinodal between the two species.

The rest of the paper is organized as follows: In Sec.\ \ref{twoc:sec}
we present the pair potentials and the full, two-component description
of the mixture, examining the effects of the depletants on the 
structural correlations of the big stars. In Sec.\ \ref{onec:sec},
we formally trace out the small stars and examine the resulting
effective, one-component interactions between the big ones. This
effective potential is employed, in turn, 
in order to draw the phase diagram of the system in Sec.\ \ref{phase:sec},
where thermodynamic perturbation theory is used for the calculation
of the Helmholtz free energies of the fluid and solid phases. 
In Sec.\ \ref{compare:sec} we carry out a comparison between the 
one- and full two-component descriptions of the mixture and demonstrate
the validity of the former, whereas in Sec.\ \ref{concl:sec} we summarize
and draw our conclusions.

\section{Two-component description}
\label{twoc:sec}

We consider binary mixtures of star-polymers which differ in 
terms of their sizes and arm numbers (functionalities). 
The system consists of $N_1$ stars of corona diameter $\sigma_1$ and
functionality $f_1$ and $N_2$ stars, 
characterized by $\sigma_2$ and $f_2$, in a volume $V$.
We first calculate the properties of the binary fluid. To obtain 
information about the pair correlations between the constituent particles,
we describe the system  using the full two-component picture 
for the mixture of the two different star-polymer species. 
The structural quantities we calculate
are used as input for the mapping onto an effective 
one-component system in Sec.\ \ref{onec:sec}.
We define the size ratio of the different species as 
$q=\sigma_2/\sigma_1 < 1$. 
Let $\rho_i=N_i/V$ ($i=1,\;2$) be the 
partial number densities of the two species. 

We start from the effective pair potentials between the mesoscopic 
particles, having traced out the monomer and solvent degrees of freedom. 
The effective interaction between
the star-polymers diverges logarithmically 
with the center-to-center distance $r$ as $r\to0$, 
as  derived by Witten and Pincus \cite{witten86}.
A full expression for identical star-polymers, 
which is valid for all star separations, 
has been derived and verified by neutron scattering and monomer resolved 
molecular simulation \cite{likos98, jusufi99}.
The pair potential is given by an ultrasoft 
interaction which shows logarithmic behavior for 
small distances and an exponential Yukawa-type
decay at large star--star separation \cite{likos98, watzlawek99}.

In the case of mixtures we need an expression for the 
effective interaction between star polymers in an athermal solvent that differ 
in their sizes $\sigma_1$, $\sigma_2$
\footnote{The $\sigma_i$ are proportional to the corresponding radii of gyration, as described in Ref.\ \cite{vonferber00}.} 
and functionalities
$f_1$, $f_2$, as a function of their center-to-center separation $r$.
In this work we use the effective pair potential which was put forward 
by means of field-theoretical arguments 
and confirmed by molecular dynamics computer 
simulations in Ref.\ \cite{vonferber00}, namely:
\begin{widetext}
\begin{equation}\label{eq.pot1}
\beta V_{ij}=\Theta_{ij}
\left\{
\begin{aligned}
&-\ln\left(\frac{r}{\sigma_{ij}}\right)+\frac{1}{1+\sigma_{ij}\kappa_{ij}}&\qquad&\text{for $r\leq\sigma_{ij}$;}\\
&\frac{1}{1+\sigma_{ij}\kappa_{ij}}\left(\frac{\sigma_{ij}}{r}\right)\exp(\sigma_{ij}\kappa_{ij}-r\kappa_{ij}) &\qquad&\text{else,}
\end{aligned}
\right.
\end{equation}
\end{widetext}
where $\sigma_{ij}=(\sigma_i+\sigma_j)/2$,  $1/\kappa_{ij}=\sigma_i/\sqrt{f_i}+\sigma_j/\sqrt{f_j}$ and
\begin{equation}
\Theta_{ij}=\frac{5}{36}\frac{1}{\sqrt{2}-1}\left[(f_i+f_j)^{3/2}-(f_i^{3/2}+f_j^{3/2})\right].
\end{equation}
Moreover, $\beta=(k_\mathrm{B}T)^{-1}$ 
is the inverse temperature, with $k_\mathrm{B}$ 
being Boltzmann's constant. Since all three interactions are purely entropic,
the $\beta V_{ij}(r)$ are independent of the temperature.
For $i=j$ the potential reduces to the interaction of 
identical star-polymers which was introduced in \cite{likos98}.
In what follows, we fix the functionality of the large stars to
$f_1 = 263$ in order to make contact with recently performed
experiments \cite{stiakakis02} in which smaller
polymeric entities were used as additives in gelated solutions of
the large stars in order to examine their overall influence on the
rheology of the mixture. This functionality is large enough for
the star polymers to crystallize into a fcc-structure roughly
at their overlap concentration \cite{watzlawek99}. For the small 
stars, we considered functionalities $f_2 = 16$ and $32$ and
size ratios $q$ in the range between $0.1$ and $0.3$.

The pair structure of the mixture can now be calculated 
using the Ornstein-Zernike (OZ) equations for binary mixtures 
together with the two-component Rogers-Young (RY)
closure. 
The pair correlations of the system are described by three 
independent total
correlation functions $h_{ij}(r)$, $i\leq j=1,2$, 
since the symmetry with respect to the exchange of indices 
dictates $h_{ij}(r)=h_{ji}(r)$ for $i \ne j$. In addition, we
have the same number of direct correlation functions $c_{ij}(r)$. 
The Fourier transforms of these quantities are denoted 
by $\tilde{h}_{ij}(k)$ and $\tilde{c}_{ij}(k)$,
respectively.

For multicomponent mixtures, the OZ relation 
takes the form \cite{lebowitz64, likos01}
\begin{equation}\label{eqOZmult}
\mathbf{\tilde{H}}(k)=\mathbf{\tilde{C}}(k)+\mathbf{\tilde{C}}(k)\cdot\mathbf{D}\cdot\mathbf{\tilde{H}}(k),
\end{equation}
where $\mathbf{\tilde{H}}(k)$ and $\mathbf{\tilde{C}}(k)$ are symmetric $\nu\times\nu$ matrices with
\begin{equation}
[\mathbf{\tilde{H}}(k)]_{ij}=\tilde{h}_{ij}(k) \qquad\text{and}\qquad [\mathbf{\tilde{C}}(k)]_{ij}=\tilde{c}_{ij}(k).
\end{equation}
$\mathbf{D}$ is a diagonal $\nu\times\nu$ matrix with
\begin{equation}
[\mathbf{D}]_{ij}=\rho_i\delta_{ij}.
\end{equation}

From Eq.\ (\ref{eqOZmult}) we obtain three independent equations 
for the six unknown functions  $\tilde{h}_{ij}(k)$ and $\tilde{c}_{ij}(k)$,
$i,j = 1,2$.
In order to obtain a solvable system, we need three additional 
{\it closure equations} between these functions. 
The Rogers-Young closure for multicomponent
mixtures reads as \cite{likos01}
\begin{equation}\label{eq.RYmult}
g_{ij}(r)=\exp[-\beta V_{ij}(r)]\left[1+\frac{\exp[\gamma_{ij}(r)f_{ij}(r)]-1}{f_{ij}(r)}\right],
\end{equation}
where $g_{ij}(r)=h_{ij}(r)+1$, $\gamma_{ij}(r)=h_{ij}(r)-c_{ij}(r)$ and $V_{ij}(r)$ is the pair interaction between particles of species $i$ and $j$. 
The ``mixing function'' $f_{ij}(r)$ is
defined as
\begin{equation}\label{eqmixing}
f_{ij}(r)=1-\exp(-\alpha_{ij}r).
\end{equation} 
Usually, the same self-consistency parameter 
$\alpha=\alpha_{ij}$ is used for all components of the mixture.
This allows to
fulfill one thermodynamic consistency requirement, 
namely the equality between the ``virial'' and ``fluctuation'' 
total compressibilities of the mixture.
Multi-parameter versions have also been proposed \cite{biben91},  
invoking thermodynamic consistency for the partial 
compressibilities of each species.
For $\alpha\to 0$ Eq.\ (\ref{eq.RYmult}) reduces to 
the Percus-Yevick (PY) and for $\alpha\to\infty$ 
to the hypernetted chain (HNC) multicomponent closures.
When dealing with star-polymers, which feature a soft repulsion of relatively short range,
neither the PY nor the HNC closure are adequate to capture the details of the correlation functions 
with high accuracy, therefore employing the full RY closure is essential \cite{watzlawek98}.

In our work we solve the OZ equation with the RY-closure 
[Eqs.\ (\ref{eqOZmult})--(\ref{eqmixing})] for the two-component mixture. 
The effective interactions 
between the star polymers are given by Eq.\ (\ref{eq.pot1}).
The thermodynamic consistency of the RY closure 
was obtained by using a single parameter $\alpha$.
The structure of the binary mixture can be described either by
the partial radial distribution functions $g_{ij}(r) = h_{ij}(r) + 1$
in real space or by the 
three partial static structure factors 
$S_{ij}(k)=\delta_{ij}+\sqrt{\rho_i\rho_j}\tilde{h}_{ij}(k)$ in
wavenumber space.
The structure factors are relevant in comparing with experiments,
because they can be measured via scattering techniques.

\begin{figure}
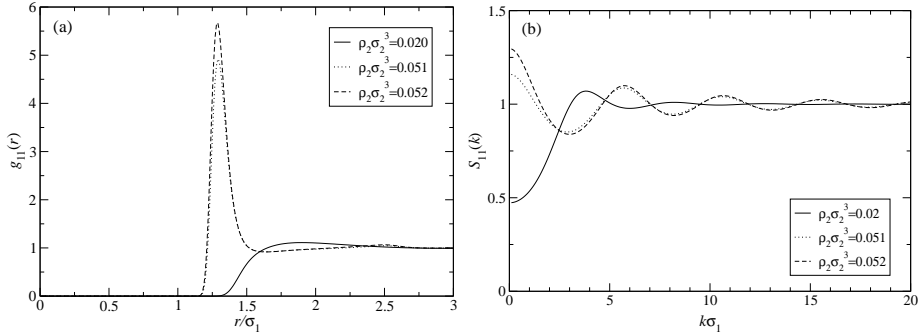

\centering
\includegraphics[width=6cm, clip=true]{fig1a.eps}
\includegraphics[width=6cm, clip=true]{fig1b.eps}
\caption{The partial (a) radial distribution functions and 
(b) static structure factors for species 1 (big stars) in a mixture
with small ones. The density of species 1 is $\rho_1\sigma_1^3=0.05$.
The structure has been calculated using the two component OZ equations with 
the RY closure. The plotted lines are for different densities $\rho_2$,
as indicated in the legend. In this case, $q=0.1$ and $f_2=32$.
The partial structure factor grows for $k\to 0$ with increasing $\rho_2$,
as we approach the demixing spinodal line of the system.}
\label{figspinodal}
\end{figure}

\begin{figure}
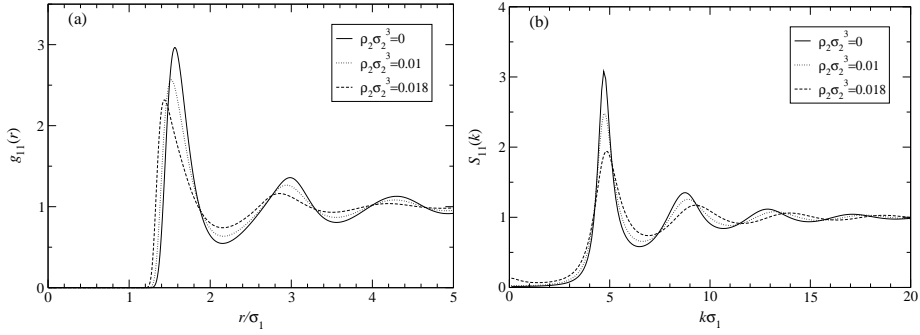

\centering
\includegraphics[width=6cm, clip=true]{fig2a.eps}
\includegraphics[width=6cm, clip=true]{fig2b.eps}
\caption{Same as Fig.\ \ref{figspinodal} but now for big star density
$\rho_1\sigma_1^3=0.3$.
By increasing 
the density of the smaller component,
the structure of the fluid diminishes.}
\label{fig.hansenverlet}
\end{figure}

In Fig.\ \ref{figspinodal}(a) we show results for the radial distribution
function $g_{11}(r)$ between the large stars in a dilute solution, and
its evolution upon increasing the concentration of small additives with
$f_2 = 32$ and $q = 0.1$. Although for very small concentration
of smaller stars the function
$g_{11}(r)$ has a relatively structureless shape, it rapidly 
develops a pronounced peak when $\rho_2$ is further increased. This is
a first indication of clustering of big stars, which has its physical
origin in some effective attraction induced by the small component.
One physically expects that when this attraction becomes sufficiently
strong, a demixing transition between the two species will take place.
This hypothesis is corroborated by the evolution of the corresponding
structure factor $S_{11}(k)$, shown in Fig.\ \ref{figspinodal}(b).
A fluid--fluid demixing binodal is indicated by
the divergence of all partial structure factors in 
the long-wavelength limit $k\to 0$. As can be seen in  
Fig.\ \ref{figspinodal}(b),
a growth of the $k \to 0$-limit occurs upon
increasing $\rho_2$. The existence of 
a demixing spinodal will be confirmed in Sec.\ \ref{phase:sec}
where we draw the phase diagram of the mixture.

We now examine the effect of the additives at the complementary
regime of high concentration of large stars, and in particular
slightly above their overlap concentration, in which the latter
are in a thermodynamically stable
crystalline state \cite{watzlawek99} or in a dynamically
arrested gel state \cite{foffi03}.
We derive the partial structure factor $S_{11}(k)$
of the (metastable) fluid in the absence of small stars and monitor
its evolution as $\rho_2$ is increased. Representative results are
shown in Fig.\ \ref{fig.hansenverlet}. In Fig.\ \ref{fig.hansenverlet}(a),
already a loss of correlations in real space can be discerned,
as witnessed by the broadening and lowering of the coordination
peaks in $g_{11}(r)$. Moreover, the large stars approach closer to
each other upon an increase of $\rho_2$, an effect that can be
interpreted as a weakening of the strength of their mutual repulsion.
As can be seen in Fig.\ \ref{fig.hansenverlet}(b),
the principle peak height of the structure factor 
of species 1 {\it diminishes} as the density of the smaller 
component is increased. The Hansen-Verlet criterion \cite{hansen69}
states that a fluid solidifies when the maximum 
of the structure factor exceeds the threshold value 
$S_{\rm th}(k_{\rm max}) = 2.85$. 
Therefore, the diminishing of structure in the
system is a first indication for the melting 
of the crystal of big star-polymers by addition 
of the smaller species. In Fig.\ \ref{fig.hansenverlet}(b)
it can be seen that the first peak of the structure 
factor is bigger than 3 for $\rho_2\sigma_2^3=0$. 
Already small densities of the smaller component
lead to a drastic decline of the peak height, a finding that is in line
with recent experimental and theoretical results on mixtures of 
star polymers with {\it linear} chains \cite{stiakakis02}.
 
\section{Effective one-component description}
\label{onec:sec}

We now wish to put the assumptions regarding the influence of the
additives on the effective interaction of the big stars into a concrete
test, by calculating an effective potential $V_{\rm eff}(r)$ between
the latter in the presence of the former. To this end,
we carry out the mapping of the two-component system 
onto an effective one-component description, in which the degrees 
of freedom of the smaller star-polymers
have been traced out.  
The interactions 
cause spatial correlations of the density of small stars in the vicinity 
of two big ones, influencing thereby the shape of the resulting
generalized depletion interaction.
There are different methods to obtain these effective interactions. 
All of them omit many-body forces and reduce the interaction 
to an additive pair potential, we will confirm however 
that many-body effects only play a minor role.

Instead of the so-called
`system representation',
in which the two densities $\rho_1$ and $\rho_2$ in the mixture are given,
we now switch into the more convenient `reservoir representation'  $\rho_1$ 
and $\rho_2^\mathrm{r}$.  
Since the effective interaction between the large stars depends rather on the 
chemical potential $\mu_2$ of the small ones rather than on
their density $\rho_2$, this description is 
more convenient \cite{md3}. The reservoir is a system consisting of pure
small stars and their density there, $\rho_2^\mathrm{r}$, is determined
by the requirement that the partial chemical potentials $\mu_2$ in the
real system and $\mu_2^\mathrm{r}$ in the reservoir are equal. Clearly, due to the 
finite value of the density $\rho_1$ in the system, it must hold
$\rho_2 \ne \rho_2^\mathrm{r}$. The mapping between the two densities,
depending parametrically on the big star density $\rho_1$, will be
carried out in Sec.\ \ref{compare:sec}.

\begin{figure}
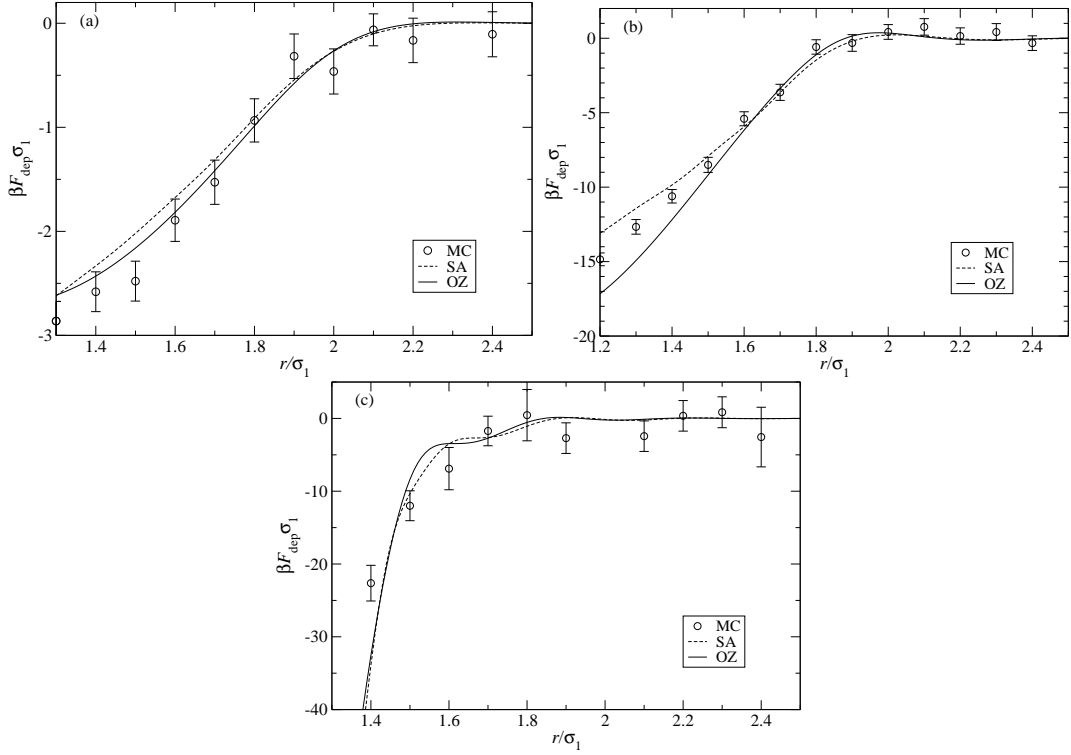

\centering
\includegraphics[width=7cm, clip=true]{fig3a.eps}
\includegraphics[width=7cm, clip=true]{fig3b.eps}
\includegraphics[width=7cm, clip=true]{fig3c.eps}
\caption{Comparison of Monte Carlo simulation (MC), 
inversion of Ornstein-Zernike equation (OZ),
and superposition approximation (SA) results for the depletion
forces between the big stars. (a) Results for $q=0.3$, $f_2=32$ 
and $\rho_2^\mathrm{r}\sigma_2^3=0.027$; (b) $q$ and $f_2$ same as in (a)
but $\rho_2^\mathrm{r}\sigma_2^3=0.081$; (c) for $q=0.2$, $f_2=32$ 
and $\rho_2^\mathrm{r}\sigma_2^3=0.08$.}
\label{fig.depl}
\end{figure}

\subsection{Simulations}\label{subsec.sim}

The most accurate way to obtain the effective interactions 
between the big star-polymers is a computer simulation 
of the mixture \cite{damico97, allahyarov98, allahyarov01}.
We place two big stars of species 1 at fixed positions  
$\mathbf{R}_1$ and $\mathbf{R}_2$ along the diagonal 
of the simulation cube, so that their common center  
coincides with the center of the cube. 
They are surrounded by the smaller species that move according to the
forces dictated by the effective interactions of Eq.\ (\ref{eq.pot1}).
Since we have only two big star-polymers in our simulation box,
the density is  $\rho_1\to 0$. Therefore the simulation provides
directly the sought-for effective force as a function of the 
reservoir density $\rho_2^\mathrm{r}$.

We use standard $NVT$ Monte Carlo simulation with periodic 
boundary conditions and minimum image convention. 
The length of the cubic simulation box 
is $L=5\sigma_1$, so that the number of small
stars in the simulation results which 
are shown in Fig.\ \ref{fig.depl} is between 125 and 1250.
For each particle up to 5 million 
Monte Carlo steps are calculated, where the maximum 
displacement of the particles is chosen
in such a way that half the steps will be accept. 
The force is then measured after every 1000 simulation steps. 
Due to the presence of the
second big star, the density distribution around 
each star is not spherically symmetric. 
This leads to an average nonvanishing force between them, 
which is mediated by the small
star-polymers and parallel to the vector 
$\mathbf{R}_{12}=\mathbf{R}_2-\mathbf{R}_1$. 
Due to the symmetry of the system, the components perpendicular 
to  $\mathbf{R}_{12}$ have to vanish. 
The force $\mathbf{F}_1$ acting on the particle 
at $\mathbf{R}_1$ can then be calculated by averaging 
over the simulation results, namely:
\begin{equation}
\mathbf{F}_1(R_{12})=\left<-\sum_{j=1}^{N_2}\nabla_{\mathbf{R}_1}V_{12}(|\mathbf{R}_1-\mathbf{r}_j|)\right>_{\mathbf{R}_{12}},
\end{equation}
where $\mathbf{r}_j$ are the star-polymer positions 
of species 2 and $\left<\cdots\right>_{\mathbf{R}_{12}}$ 
denotes the statistical average, taken under the constraint
of constant $\mathbf{R}_{12}$. Clearly, the effective force
satisfies the relation $\mathbf{F}_2(R_{12})=-\mathbf{F}_1(R_{12})$. 
We further define the {\it depletion} force ${\bf F}_\mathrm{dep}(R_{12})$
as the difference between $\mathbf{F}_2(R_{12})$ and the direct 
force ${\bf F}_{\rm {dir}}(R_{12})$ between the two stars due to their
direct interaction potential $V_{11}(R_{12})$.
The magnitude of the depletion force $F_\mathrm{dep}(R_{12})$ is 
then given by
\begin{equation}
F_\mathrm{dep}(R_{12})=
\frac{\mathbf{R}_{12}}{R_{12}}\cdot\mathbf{F}_{\rm dep}(R_{12}).
\end{equation}
Accordingly,
the total effective interaction between the big star-polymers 
in a sea of the smaller species is the sum of their 
interaction potential $V_{11}(r)$ and the depletion potential 
$V_\mathrm{dep}(r)$:
\begin{equation}
V_{\rm eff}(r) = V_{11}(r) + V_\mathrm{dep}(r).
\label{vdep:eq}
\end{equation}

A large number of long simulation runs is required to 
to obtain accurate depletion forces with good statistics, which renders
this approach inefficient if ones needs to calculate $V_{\rm eff}(r)$
for arbitrary values of $q$, $f_2$ and $\rho_2^{\rm r}$. Thus, we 
resorted to 
approximative theoretical methods to calculate
the effective interactions and used 
the simulation results at selected parameter combinations in order to
put the theoretical approximations into test. The two theoretical
approaches invoked in this work are the inversion of the 
Ornstein-Zernike equation and the superposition approximation,
which are presented below.

\subsection{Inversion of the Ornstein-Zernike equation}\label{subsecinvOZ}

The effective potential can be obtained by inversion 
of the two-component OZ equation results in the 
limit of low density of big stars \cite{dzubiella02, mendez00, konig01}.
It can be shown from diagrammatic expansions in the theory of 
liquids \cite{hansen86} that the radial distribution 
function $g(r)$ of any fluid whose constituent particles
interact via the pair potential $V(r)$, reduces to the Boltzmann factor
$g(r)=\exp[-\beta V(r)]$ in the low-density limit. 
The effective interaction between the big stars 
depends on the reservoir density $\rho_2^{\rm r}$ of the 
smaller component. The interaction can be obtained by 
solving the full two-component OZ equations with the RY closure 
for different small-component densities $\rho_2$
in the limit $\rho_1 \to 0$; due to the latter limit, it then also
holds $\rho_2 = \rho_2^{\rm r}$.
The radial distribution function $g_{11}(r)$ can then be 
inverted to yield the effective potential as
\begin{equation}
\beta V_\mathrm{eff}(r)=
-\lim_{\rho_1\to 0}\ln[g_{11}(r;\rho_1,\rho_2^\mathrm{r})].
\end{equation}
Thereafter, the depletion force $F_\mathrm{dep}(r)$ can be calculated as 
$F_\mathrm{dep}(r)=-\partial[V_\mathrm{eff}(r)-V_{11}(r)]/{\partial r}$ 
and compared them to the simulation results of the preceding subsection.
Selected comparisons are shown in 
Fig.\ \ref{fig.depl}, where it can be seen that the inversion of the
OZ relation yields very reliable results. We emphasize here that the
approximate character of the OZ-inversion technique lies exclusively
in the approximations involved in solving the two-component integral
equation theories, i.e., in the Rogers-Young (or any other chosen)
closure relation. Otherwise, the method is based on the exact statement
that the radial distribution function of a one-component system at
low densities is equal to the Boltzmann factor of the associated
pair potential and hence the agreement of the inversion method with
the simulation results comes as no surprise. It rather corroborates
the fact that the two-component RY-closure is very accurate whenever
one deals with soft, repulsive interactions, a result already seen
in the case of mixtures between hard spheres and 
star polymers \cite{dzubiella02a}.

\subsection{Superposition approximation}

Another possibility to derive the effective 
interaction is the \textit{superposition approximation} 
(SA) \cite{attard89}. If the density distribution  
$\rho_2(\mathbf{r}; \mathbf{R}_1,\mathbf{R}_2)$ of the small stars
around two big stars held fixed at positions ${\bf R}_1$ and ${\bf R}_2$
is known, then the depletion force
in the low-density limit can be calculated by a simple integration.
The density $\rho_2(\mathbf{r}; \mathbf{R}_1,\mathbf{R}_2)$ 
is proportional to three-body distribution function 
$g_{112}(\mathbf{R}_1, \mathbf{R}_2, \mathbf{r})$,
which expresses the probability density of finding a particle of
species 2 at position ${\bf r}$, given that two particles of
species one are fixed at positions ${\bf R}_1$ and
${\bf R}_2$. This 
function is in general unknown; a usual procedure is to 
approximate it by the product of pair distribution 
functions \cite{attard89}.

We consider two big stars at the positions  
$\mathbf{R}_1$ and  $\mathbf{R}_2$ and 
choose, without loss of generality, $\mathbf{R}_1=0$. 
Let the distance between the particles be $R_{12}$.
The surrounding smaller star-polymers have the 
density $\rho_2(\mathbf{r}; \mathbf{R}_1,\mathbf{R}_2)$. 
By taking the average for fixed $R_{12}$ we obtain the depletion force as
\begin{eqnarray}
F_\mathrm{dep}(R_{12})&=&-2\pi\int_0^\infty r^2\frac{\mathrm{d}V_{12}(r)}{\mathrm{d}r}\mathrm{d}r \nonumber \\
&&\int_{-1}^1\rho_2(\mathbf{r}; \mathbf{R}_1,\mathbf{R}_2)\omega \mathrm{d}\omega,
\end{eqnarray}
where  $\omega=\cos\theta$. 

\begin{figure}[h]
\includegraphics[width=6cm, clip=true]{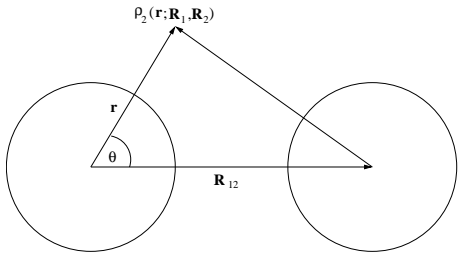}
\caption{A sketch of two big stars 
at a distance $R_{12}$ with 
$\rho_2(\mathbf{r}; \mathbf{R}_1, \mathbf{R}_2)$ 
denoting the density of the smaller stars at $\mathbf{r}$. 
The density distribution 
depends on
the positions of the two big star-polymers.}
\end{figure}

Since  $\rho_2(\mathbf{r}; \mathbf{R}_1,\mathbf{R}_2)$ 
is in general not known, at this point the exact 
density distribution has to be approximated. The density distribution
around two big stars is replaced by the product of the 
density distributions around two isolated star-polymers 
at the positions  $\mathbf{R}_1$ and  $\mathbf{R}_2$,
respectively. The SA then reads as
\begin{equation}
 \rho_2(\mathbf{r}; \mathbf{R}_1,\mathbf{R}_2)\approx\rho_2^\mathrm{r}g_{12}(|\mathbf{r}-\mathbf{R}_1|)g_{12}(|\mathbf{R}_2-\mathbf{r}|),
\end{equation}
where $\rho_2^\mathrm{r}$ is the reservoir density, again identical to
the system density for the situation at hand, since only two big stars
are considered in the thermodynamic limit and thus $\rho_1 = 0$.
The functions $g_{12}(|\mathbf{R}_i-\mathbf{r}|)$ 
are the radial distribution functions of small star polymers surrounding a
single large one.
Therefore, they can be obtained in the $\rho_1 \to 0$-limit 
of the two-component OZ equations. 
Using simple geometrical considerations,
we obtain $|\mathbf{r}-\mathbf{R}_1|=\sqrt{R_{12}^2+r^2-R_{12}r\omega}$. 
Finally we obtain for the depletion force in the SA the expression:
\begin{eqnarray}
\nonumber
F_\mathrm{dep}(R_{12})&=&-2\pi\rho_2^\mathrm{r}\int_0^\infty r^2\frac{\mathrm{d}V_{12}(r)}{\mathrm{d}r}g_{12}(r)\mathrm{d}r
\\
\nonumber
&\times& \int_{-1}^1g_{12}\left(\sqrt{R_{12}^2+r^2-R_{12}r\omega}\right)\omega \mathrm{d}\omega.
\\
& &
\end{eqnarray}

The results we obtain by this method are compared 
to Monte Carlo and to those derived from inversion of the 
Ornstein-Zernike equation in Fig.\ \ref{fig.depl}. The results are
very similar to the ones we obtain by inverting the OZ equation 
and both approximations yields reasonable agreement with the 
simulation data. Therefore, it is
possible to choose the results of either approximation 
for calculating the phase diagrams and we expect that only minor 
quantitative differences will be seen by employing the one or the
other theoretical approach. We have chosen to work with the effective
interactions that result from the inversion of the OZ-equation since
the latter is based on an exact statement, whereas the SA has an
approximate nature.

\subsection{Effective interactions}\label{subsec.effint}

\begin{figure*}
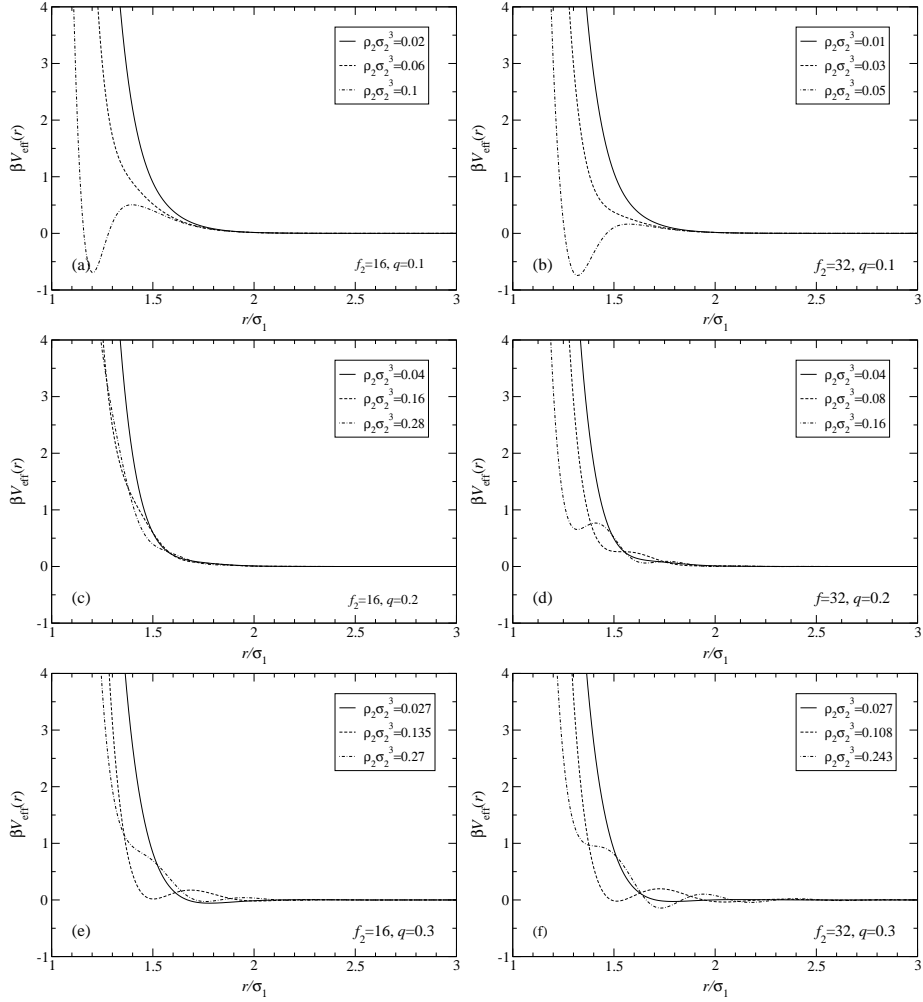

\begin{center}
\includegraphics[width=6cm, clip=true]{fig5a.eps}
\includegraphics[width=6cm, clip=true]{fig5b.eps}
\includegraphics[width=6cm, clip=true]{fig5c.eps}
\includegraphics[width=6cm, clip=true]{fig5d.eps}
\includegraphics[width=6cm, clip=true]{fig5e.eps}
\includegraphics[width=6cm, clip=true]{fig5f.eps}
\end{center}
\caption{The effective potential $V_\mathrm{eff}(r)$ between two big
star polymers in the presence of a sea of small ones. The various
combinations of parameters regarding the density, functionality and
size of the additives are shown in the legends. Notice the development
of a strong attractive part in the interaction for the case $q = 0.1$.}
\label{figveff}
\end{figure*}

We have chosen to employ the effective
interactions that result from the inversion of the OZ-equation since
the latter is based on an exact statement, whereas the SA has an
approximate nature.
Representative results are shown
in Fig.\ \ref{figveff}. For the lowest size ratio, $q = 0.1$,
we see that irrespective of the functionality of the depletants
($f_2 = 16$ or $32$), the following scenario materializes: as 
$\rho_2^{\rm r}$ increases, first a weakening of the repulsions takes
place, followed by the development of an attraction between the
stars at sufficiently high reservoir densities, see
Figs.\ \ref{figveff}(a) and \ref{figveff}(b). These findings
provide a possible physical realization of the recently-proposed
model ultrasoft repulsion potentials that are accompanied by 
an attractive part \cite{loverso:03}.
This attraction is more pronounced for $f_2=32$ than for $f_2=16$,
if one compares two systems with equal
density $\rho_2^\mathrm{r}\sigma_2^3$. This result is not surprising,
since the $f_2 = 32$-stars exert a higher osmotic pressure on the
large ones than the $f_2 = 16$-stars and can therefore reduce the
direct repulsions and induce attractions more efficiently.

Novel features in the effective potential appear for higher size ratios,
$q = 0.2$ and $q = 0.3$. As can be seen in Fig.\ \ref{figveff}(c)-(f),
an oscillatory structure appears in the effective potential, which
is akin to that seen for hard-sphere mixtures of two different sizes.
Contrary to this case, however, a deep attraction between the big
stars does not develop and, therefore, it seems that a demixing 
transition between the two species does not exist when the sizes
of the two stars become more and more similar. 
In all cases, however, the range of the repulsion decreases 
due to the depletion effect, i.e., the 
big star polymers appear, in the presence of the small ones,
to be softer than they are in a pure solvent. Another possible interpretation,
to be elaborated on in what follows, is that the big stars appear to be
`smaller', i.e., they acquire a reduced  
effective hard sphere packing fraction as a result of the depletants.
Since the star-polymers then need less space, they
become more mobile so the solid can melt. 
This property will be discussed 
in more detail in Sec.\ \ref{subsec.phaseresults}.

\section{Phase diagrams}
\label{phase:sec}

\subsection{Hard sphere mapping}

\begin{figure}
\centering
\includegraphics[width=7.2cm, clip=true]{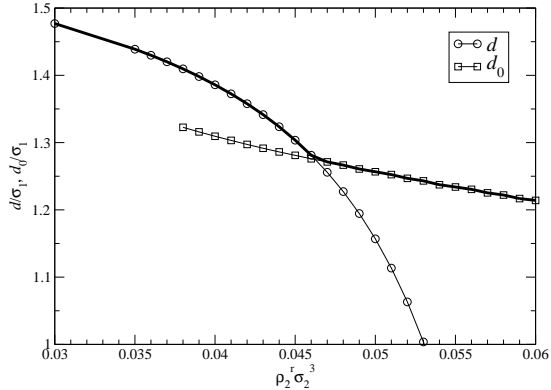}
\caption{The two possible effective hard sphere diameters $d$ and $d_0$
pertaining to the effective interaction $V_{\rm eff}(r)$ between big
stars, against the reservoir density $\rho_2^{\rm r}$. The parameter
combination here is $f_2=32$ and $q=0.1$. 
As explained in the text, $d$ is calculated using the full 
effective interaction and $d_0$ only for the reference part.}
\label{fig.diameter}
\end{figure}

In order to trace out the phase diagram of the mixture in the
$(\rho_1, \rho_2^{\rm r})$-representation, we first perform a
mapping of the effective one-component interaction $V_{\rm eff}(r)$
between the big stars onto an effective hard-sphere system of
diameter $\sigma$. Clearly, the latter depends on the reservoir density
$\rho_2^{\rm r}$ as well as on the system parameters $q$ and $f_2$.
For the purposes of performing the mapping in a physically meaningful
way, we distinguish between two cases. 

First, we consider the case in which $V_{\rm eff}(r)$ is either
free of attractive parts or positive definite or, at most, it
contains negative parts not exceeding a small fraction of $k_{\rm B}T$
in magnitude. In this case, it is physically meaningful to identify 
$\sigma$ with the Barker-Henderson hard sphere diameter of the
{\it full} effective interaction $V_{\rm eff}(r)$, $d$, defined
as \cite{barker67}:
\begin{equation}
d=\int_0^\infty\left\{1-\exp[-\beta V_{\rm eff}(r)]\right\}\mathrm{d}r.
\label{eq.barkerhend}
\end{equation}
Most of the curves shown in Fig.\ \ref{figveff} fall into this category.
An important exception are the curves pertaining to 
$\rho_2^{\rm r}\sigma_2^3 = 0.1$ in Fig.\ \ref{figveff}(a) and
to $\rho_2^{\rm r}\sigma_2^3 = 0.05$ in Fig.\ \ref{figveff}(b). For
these combinations, and also for all others at even higher 
reservoir densities, a deep negative minimum appears in $V_{\rm eff}(r)$
and application of Eq.\ (\ref{eq.barkerhend}) to such cases would
lead to unphysically small and even negative effective hard sphere
diameters. For such combinations, it is physically appealing to
separate the effective potential $V_{\rm eff}(r)$ into a purely
repulsive part $V_0(r)$ and a perturbation part $V_{\rm {pert}}(r)$,
by truncating and shifting upwards the full interaction at the
deepest minimum \cite{barker67}. In this second case, it is pertinent
to define another effective hard sphere diameter, $d_0$, that is
associated with $V_0(r)$ only and is calculated again from the
Barker-Henderson recipe, namely
\begin{equation}
d_0=\int_0^\infty\left\{1-\exp[-\beta V_0(r)]\right\}\mathrm{d}r.
\label{eq.barkerhend0}
\end{equation}

In attempting to choose and match between the two possible hard sphere
diameters, $d$ and $d_0$, we are confronted with a technical difficulty.
The evolution of the potential $V_{\rm eff}(r)$ with $\rho_2^{\rm r}$
is continuous and the appearance of negative minima is in general
accompanied by a soft repulsive barrier after the minimum. The effective
hard sphere diameter, on the other hand, has to be a continuous 
function of $\rho_2^{\rm r}$, so as to avoid unphysical jumps of the
phase boundaries in the phase diagram. In Fig.\ \ref{fig.diameter}
we show a typical result for the dependence of $d$ and $d_0$ 
on $\rho_2^{\rm r}$. For low values of $\rho_2^{\rm r}$, where
$V_{\rm eff}(r)$ is purely repulsive, $d$ is a meaningful measure
of the effective hard-sphere diameter. On the other hand, at high
values of $\rho_2^{\rm r}$, where a deep attraction between the big stars
effectively sets in, it is $d_0$ that most realistically captures
the physics of the repulsions. The two curves cross at some point
and, in order to guarantee both the continuity of $\sigma$ as a
function of $\rho_2^{\rm r}$ and its correct asymptotic behavior for
small and large values of $\rho_2^{\rm r}$, we choose
\begin{equation}
\sigma = \max\{d,d_0\}.
\label{sigma:eq}
\end{equation}
It is then clear from the discussion above that for the perturbation
part, $V_{\rm {pert}}(r)$, of the interaction, it holds
\begin{equation}
V_{\rm {pert}}(r) = 0 \qquad{\rm {if}\,} d > d_0.
\label{pert:eq}
\end{equation}
 
The phase diagrams can now be calculated using standard 
first-order perturbation theory \cite{hansen86} and taking
Eqs.\ (\ref{sigma:eq}) and (\ref{pert:eq}) into account. 
We do not take higher orders into account because we
are mainly interested in the qualitative behaviour of the freezing line
for small densities  $\rho_2^\mathrm{r}$.
Denoting by $F_0$ the Helmholtz free energy of the reference 
hard sphere system (effective hard sphere diameter $\sigma$), 
the total Helmholtz free energy $F$ of the one-component system
consisting of $N_1$ big star polymers is approximated by
\begin{equation}
\frac{\beta F}{N_1}=\frac{\beta F_0}{N_1}
+\frac{1}{2}\beta\rho_1\int g_0(r)V_\mathrm{pert}(r)\mathrm{d}^3r.
\label{f0:eq}
\end{equation} 
In Eq.\ (\ref{f0:eq}) above, $g_0(r)$ denotes the radial distribution
function of the reference hard-sphere system in the fluid phase 
and its angle-averaged counterpart in the solid phase, as
defined in Ref.\ \cite{kincaid77}.

We note here that 
more accurate methods for the treatment of 
potentials with a soft cores have also been proposed \cite{ben03}, 
but are not used here since we are only interested in
the basic topology of the phase diagrams. 
For the free energy of the reference hard sphere system we used   
the equations of state  of the Carnahan-Starling \cite{carnahan69}
and Hall \cite{hall70} for the solid and fluid phases, respectively. 
For the calculation of pair distribution 
functions we use the expressions of Henderson and Grundke
\cite{henderson75} for the fluid and 
of Kincaid and Weis \cite{kincaid77} for the fcc-solid. 
We only considered the fcc-solid because this crystal structure
appears at the fluid--solid transition in one component 
star-polymer solutions with an arm number $f_1 =  263$ \cite{watzlawek99}. 

\subsection{Results}\label{subsec.phaseresults}

\begin{figure*}
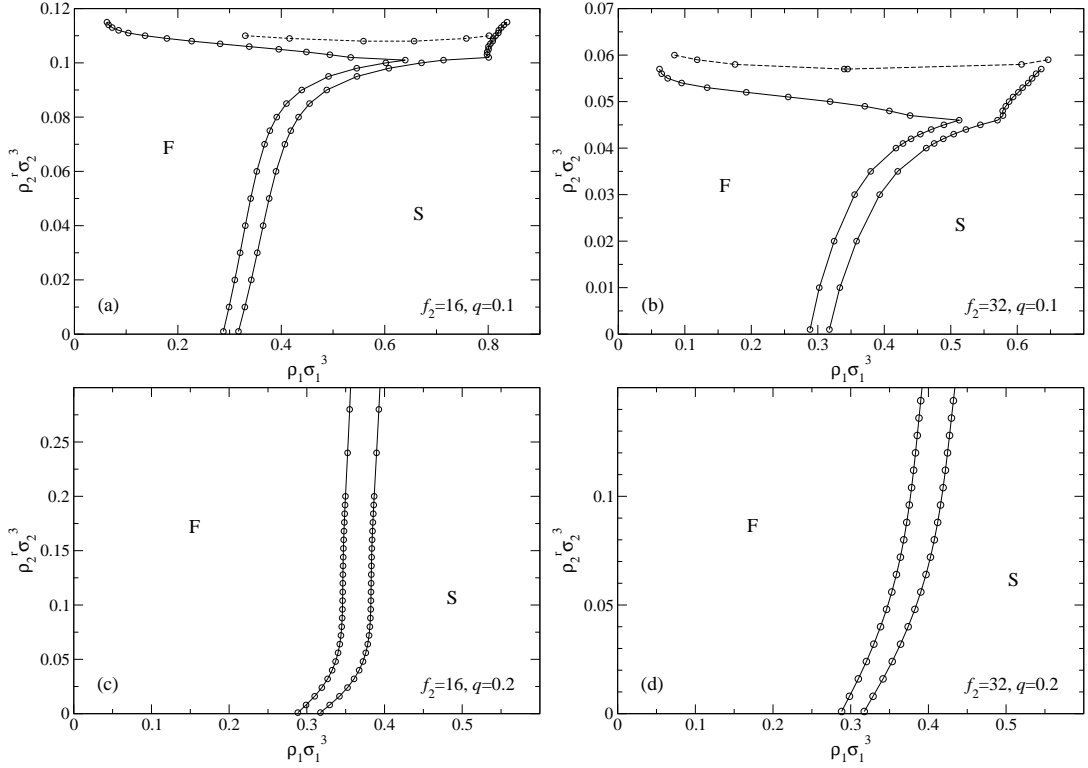

\centering
\includegraphics[width=7.1cm, clip=true]{fig7a.eps}
\includegraphics[width=7.1cm, clip=true]{fig7b.eps}
\includegraphics[width=7.1cm, clip=true]{fig7c.eps}
\includegraphics[width=7.1cm, clip=true]{fig7d.eps}
\caption{Phase diagrams for star polymer mixtures 
for different parameter combinations. The big star functionality is
fixed at $f_1 = 263$. 
The circles denote the phase boundaries as calculated by the hard-sphere
mapping including the perturbation part $V_{\rm {pert}}(r)$ and the 
lines are a guide to the eye. 
It can be seen that we only find a fluid--fluid demixing for $q=0.1$. 
This binodal line is metastable with respect to the formation
of the fcc-solid. The kink in the phase boundaries for $q=0.1$ 
is an artifact and is caused by the method used 
to split the effective pair potential
into reference and perturbation part. The symbol F stands for the fluid
and the symbol S for the solid regions.}
\label{figphases}
\end{figure*}

In Fig.\ \ref{figphases} we show
the resulting phase diagrams for size ratios 
$q=0.1$ and $q=0.2$ and 
functionalities of the smaller species $f_2=16$ and $f_2=32$, as obtained
by the procedure described in the preceding subsection.
The kinks for $q=0.1$ at about $\rho_1\sigma_1^3=0.8$ 
and $\rho_2\sigma_2^3=0.1$ and 
$\rho_1\sigma_1^3=0.6$ and $\rho_2\sigma_2^3=0.05$, respectively,  
are an artifact of the choice (\ref{sigma:eq}) and are associated with
the sudden appearance of the $V_{\rm {pert}}(r)$-term in 
Eq.\ (\ref{f0:eq}), once we cross over from the case
$d > d_0$ to the case $d < d_0$ (cf.\ also Fig.\ \ref{fig.diameter}).
Since we are primarily interested 
in the behavior for small densities $\rho_2^\mathrm{r}$ and the
influence of the additives on crystallization, on the one hand,
and on the possible {\it existence} of a spinodal line, on the other,
a more sophisticated approach to the problem is at this stage not
necessary. The liquid--solid coexistence region 
obtained by this approach is rather wide, due to the mapping on the
effective hard-sphere system. Accurate calculations of the phase
diagram of star polymers reveal that 
the coexistence region is much more narrow \cite{watzlawek99}, yet
the shape and evolution of the freezing lines as a function of
$\rho_2^{\rm r}$ are not influenced by the width of the density gap
between the fluid and the solid phases. Finally, we note that
we have shifted the freezing line to higher densities by 
an amount $\Delta\rho_1\sigma_1^3=0.06$, in order  
to obtain the same density values for the crystallization 
as in the accurately known one component case \cite{watzlawek99}.

For size ratio $q=0.1$, 
the effective potential $V_\mathrm{eff}(r)$ develops a strong attraction. 
This leads to a broadening of the coexistence area between the solid
and fluid phase and eventually to a demixing binodal. 
This binodal is found, however, to be metastable 
with respect to the crystallization. The sharp kinks that show
up in Figs.\ \ref{figphases}(a) and \ref{figphases}(b) are
artifacts of the way in which the effective hard sphere diameter
$\sigma$ was determined and, in particular, of the fact that 
the attractive perturbation part $V_{\rm pert}(r)$ of the effective
potential is absent in the treatment for small reservoir densities
below the kink and present above it. In reality, we expect the 
phase coexistence lines to `turn around' smoothly, i.e., without
the aforementioned artificial kink. However, the topology of the
phase diagram and in particular the positive slope of the freezing
lines and the subsequent broadening of the coexistence region into
a `gas-crystal' phase separation is not expected to be affected by
these approximations.
For larger values of $q$,
a strong attraction
does not emerge, so there is no fluid-fluid demixing. 
From  Fig.\ \ref{figphases} it can be seen that 
less stars with $f_2=32$ than with $f_2=16$ are needed to achieve 
similar  effect. This is in agreement with the 
properties of star-polymer--colloid mixtures 
which were investigated in \cite{dzubiella02a}.
The metastable binodal is closer to the stable region 
of the phase diagram for the smaller functionality of the 
depletant. The same trends were also observed in star-polymer--colloid
mixtures \cite{dzubiella02a},
where stable binodals were only found for small 
depletant functionalities such as 
$f_2 = 2$ and $f_2 = 6$. 
For the star-polymer mixtures we consider here, the existence of a
binodal is less likely than in star-polymer--colloid mixtures, 
because for star polymers the depletion force 
has to overcome the Yukawa-type repulsion between them before an
effective attraction sets in.

A striking effect is the melting of the crystal of the big stars upon
addition of the small component, as can be seen from the positive
slope of the freezing- and melting lines in Fig.\ \ref{figphases}.
No effective attraction between the star-polymers is needed for this 
effect.
As can be seen from Fig.\ \ref{fig.diameter}, the effective hard sphere
diameter $\sigma$ of the system decreases with increasing depletant
density. Therefore, the effective packing fraction 
$\eta_{\rm {HS}} = (\pi/6)\rho_1\sigma^3$ for the big stars gets smaller 
with increasing $\rho_2^{\rm r}$. 
Since this 
quantity determines the location of the freezing transition, 
the latter shifts to higher values of $\rho_1$.
The melting is therefore caused by the 
apparent shrinkage of the stars due
to the depletion effects.
The counterintuitive behavior of melting of a crystal although the
overall polymer concentration is increased has its physical roots in
the {\it soft depletion} mechanism and the associated reduction of
the range of the repulsive potential between the big stars. 

In real experimental systems of star polymers,  
usually the formation of a glass is observed 
instead of crystallization. These glass transition lines are usually parallel
to the freezing lines of the phase diagram, a 
property explicitly confirmed for  
the glass line of one-component star polymer solutions \cite{foffi03}. 
Therefore, we expect that the kinetic phase diagram of the mixture will
have a topology running parallel to that of the equilibrium one that
we traced in this study. We predict, therefore, that addition of small
stars will bring about a melting of the colloidal glass (or gel)
formed by the large ones. This has been shown to be the case for 
mixtures of star polymers with linear homopolymer chains \cite{stiakakis02}. 

\section{Comparison between one- and two-component descriptions} 
\label{compare:sec}

\subsection{Chemical potentials}

By calculating the partial chemical potentials 
of the two-component system, we now map the reservoir 
representation $(\rho_1, \rho_2^\mathrm{r})$ on the real
physical system $(\rho_1, \rho_2)$, so as to make contact with
experimental work, in which $\rho_2^{\rm r}$ has no direct relevance. 
The densities $\rho_2^\mathrm{r}$ and $\rho_2$ 
are linked by the condition that the chemical potential of the reservoir
$\mu_2^\mathrm{r}$ has to be the same as the 
partial chemical potential in the system $\mu_2$. 

The chemical potential in the reservoir 
of density $\rho_2^{\rm r}$ can easily be calculated 
if the fluid structure is known. The Helmholtz free energy density 
of the small stars in the reservoir,
$F/V = f(\rho_2^\mathrm{r})$, can be split into an ideal and an excess part:
\begin{equation}
f(\rho_2^\mathrm{r})=f_\mathrm{id}(\rho_2^\mathrm{r})
                     +f_\mathrm{ex}(\rho_2^\mathrm{r}),
\end{equation}
where $f_\mathrm{id}(z)=\rho\ln(z\tau^3)-z$ and $\tau$ is an arbitrary
length scale.
The second derivative of the excess 
part is connected to the structure via the relation \cite{likos95}
\begin{equation}
f_\mathrm{ex}^{\prime\prime}(\rho_2^\mathrm{r})
=\frac{1}{\rho_2^\mathrm{r} S(k=0;\rho_2^\mathrm{r})}
-\frac{1}{\rho_2^\mathrm{r}},
\end{equation}
where $S(k)$ is the structure factor of the small stars 
in the reservoir (one-component system).
The free energy can then be calculated by two integrations, with
the integration constants determined 
by the conditions $f_\mathrm{ex}^\prime(0)=0$ and
$f_\mathrm{ex}(0)=0$. The chemical potential of the reservoir is given by
\begin{equation}\label{chempot.eq}
\mu_2^\mathrm{r}=f^\prime(\rho_2^\mathrm{r}).
\end{equation} 

In order two calculate the partial chemical potential $\mu_2$ 
in the two-component system representation, 
we employ the so-called concentration structure factor \cite{bhatia70}
\begin{equation}
S_\mathrm{con}(k)=x_1x_2^2S_{11}(k)+x_1^2x_2S_{22}(k)-(x_1x_2)^{3/2}S_{12}(k),
\end{equation} 
where $x_i=\rho_i/(\rho_1+\rho_2)$. Thermodynamic properties 
can then be calculated by using the equation \cite{biben91a, bhatia70}
\begin{equation}\label{gibbs.eq}
\lim_{k\to 0}S_\mathrm{con}(k)=\left[\frac{\partial^2 g(x_2, P, T)}{\partial x_2^2}\right]^{-1},
\end{equation}
where $g(x_2, P, T)=G(x_2, N, P, T)/N$ is 
the Gibbs free energy per particle of the 
two-component mixture and $P$ its total pressure. 
Eq.\ (\ref{gibbs.eq}) can then be integrated as 
described in Ref.\ \cite{dzubiella02a} 
to yield the sought-for Gibbs free energy per particle $g(x_2, P, T)$. 
Once the Gibbs free energy is known, 
the partial chemical potentials can be calculated using the equations
\begin{equation}\label{chempot1.eq}
g^\prime(x_2)=\mu_2-\mu_1
\end{equation} 
and
\begin{equation}\label{chempot2.eq}
g(x_2)-x_2g^\prime(x_2)=\mu_1.
\end{equation}
Reverting from the pair of variables $(x_2, P)$ 
back to $(\rho_1, \rho_2)$ for the mixture
and using Eqs.\ref{chempot.eq}, \ref{chempot1.eq} and \ref{chempot2.eq}, 
the mapping 
$\rho_2^\mathrm{r}) \rightarrow (\rho_1, \rho_2)$ 
can be carried out.
\begin{figure}
\centering
\includegraphics[width=6cm, clip=true]{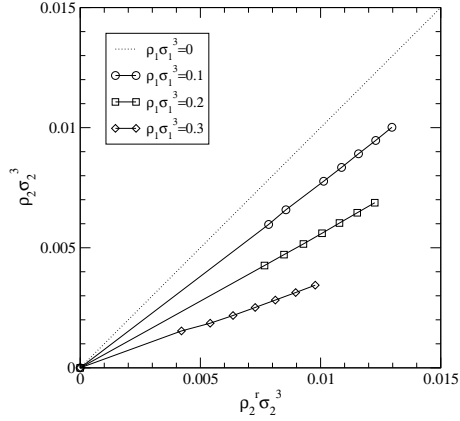}
\caption{Mapping between the system- and reservoir densities 
$\rho_2$ and $\rho_2^{\rm r}$ of the small stars
for $f_2=32$ and $q=0.1$, and for various different values of the
big star density $\rho_1$.}
\label{fig.mapping}
\end{figure}
Representative results are shown in Fig.\ \ref{fig.mapping}.
Clearly, the mapping depends parametrically 
on the big star density $\rho_1$ in the mixture.
For $\rho_1=0$ one recovers $\rho_2=\rho_2^\mathrm{r}$.
In all cases, we obtain $\rho_2<\rho_s^\mathrm{r}$ 
because all interactions are purely repulsive. 
The difference between the reservoir and system densities grows with
increasing $\rho_1$.

\subsection{Structure}

\begin{figure*}
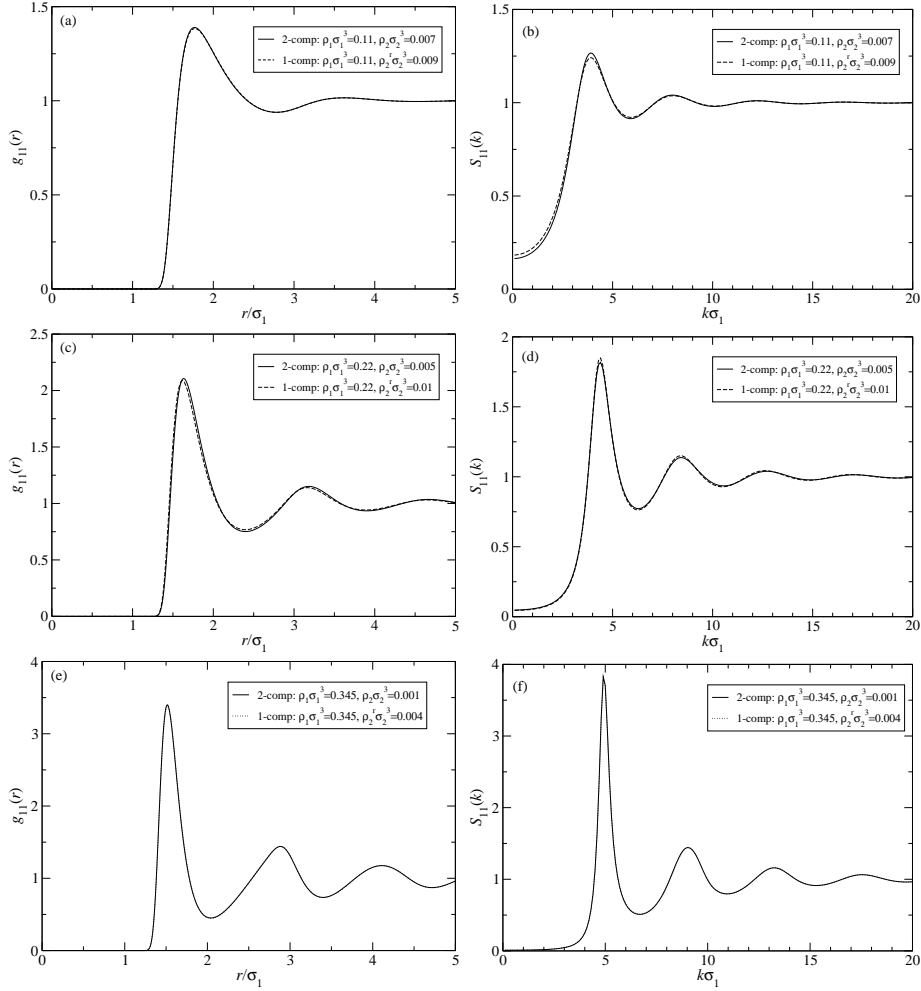

\centering
\includegraphics[width=6cm, clip=true]{fig9a.eps}
\includegraphics[width=6cm, clip=true]{fig9b.eps}
\includegraphics[width=6cm, clip=true]{fig9c.eps}
\includegraphics[width=6cm, clip=true]{fig9d.eps}
\includegraphics[width=6cm, clip=true]{fig9e.eps}
\includegraphics[width=6cm, clip=true]{fig9f.eps}
\caption{Comparison of the fluid structure in the 
two- and effective one-component case. 
The densities are denoted in the diagrams. 
The structure of the mixture
was calculated using the two-component RY closure, 
in the one component the usual RY closure was used. 
All plots are for $q=0.1$ and $f_2=32$. In panels (a), (e) and (f) 
the solid and dashed lines fall on top of each other, so that the
latter cannot be distinguished from the former.}
\label{figcompare}
\end{figure*}

We now consider the spatial correlations between the 
big stars in the fluid phase. We compare the 
correlation functions calculated in the two-component 
description with those arising from 
the one-component description using the effective 
interactions in the presence of the smaller species. 
The comparison has to be carried out for parameter 
combinations such that the reservoir- and system
partial chemical potentials of the smaller species
are equal to one another.  
As the one-component description reduces the effective 
interaction to pair potentials, the comparison of 
the structural properties allows to estimate the magnitude of
many body effects on the depletion.

In 
Fig.\ \ref{figcompare} we show the radial distribution 
functions and static structure factors of the big stars 
calculated with both different approaches.
For the one-component case, the RY closure can be used 
because for the densities we consider here 
the effective interactions remain purely repulsive. 
The cases we consider here 
are for small densities $\rho_2$ but a wide range of values of $\rho_1$. 
The two component description includes many-body forces between the 
big stars which are caused by the smaller component. 
These effects are neglected
in the effective one-component description \cite{md3}.
In Fig.\  \ref{figcompare}, one can see excellent agreement 
of the results obtained by using these two 
different descriptions of the physical system. This additionally
corroborates the 
validity of our approximation of the effective interaction. It can be 
also be concluded from the plots that the three-body forces 
are indeed much weaker than the pair interactions and can be
safely neglected \cite{melchionna00, goulding01}.

\section{Summary and conclusions}
\label{concl:sec}

We have analyzed the structural and phase behavior of highly asymmetric 
mixtures of star polymers, with the asymmetry characterizing both their
sizes and functionalities. The most striking phenomenon predicted
by our investigations is the counter-intuitive {\it melting} of the
colloidal crystal of the big stars upon addition of small ones. Though
this finding appears paradoxical at first sight, since addition of
smaller stars increases the overall polymer concentration of the 
solution, its physical explanation can be traced to the effects of
{\it soft depletion}. Whereas the depletants of hard spheres simply
superimpose an effective attraction on a hard potential, when the
big particles are themselves repulsive the depletion attraction
is superimposed on a (soft) repulsion. In this way, a repulsive
potential of reduced strength and/or range results and the effective,
reduced repulsion is not any more sufficient to maintain the stability
of the crystal, which therefore melts. When the depletant concentration
becomes sufficiently high, the attractive depletion force dominates
over the soft repulsion, leading to a (possibly metastable) demixing
transition between the two species.

In real experimental systems of star polymers with high functionality,
crystallization is hindered by the vitrification (gelation) transition
and the large stars become structurally arrested in a glassy state
above the overlap concentration. The next step would be then to 
investigate the role and influence of smaller star additives on 
the glass transition of larger stars, a problem of significant 
importance for the control of the rheological behavior of soft matter
through additives. Work along these lines is currently under way
and the presentation of the results of this investigation will be
the subject of a future publication.

\begin{acknowledgments}
We thank Dimitris Vlassopoulos, Joachim Dzubiella, Martin Konieczny, 
Emanuela Zaccarelli, Francesco Sciortino, and
Piero Tartaglia for helpful discussions. This work has been supported
by the Marie Curie European Network MRTN-CT-2003-504712 and by the DFG within SFB TR 6.
\end{acknowledgments}

\end{document}